\documentstyle[multicol,prl,aps,psfig]{revtex}

\begin{document}

\draft

\title{ Nonequilibrium critical dynamics of a three species
monomer-monomer model}

\author{Kevin E.\ Bassler and Dana A. Browne}

\address{
Department of Physics and Astronomy, Louisiana State University, Baton
Rouge, LA 70803 }

\date{June 11, 1996; revised August 28, 1996}
\maketitle{}

\begin{abstract}

We study a three species monomer-monomer catalytic surface reaction
model with a reactive steady state bordered by three equivalent
unreactive phases where the surface is saturated with one species.  The
transition from the reactive to a saturated phase shows directed
percolation critical behavior.  Each pair of these reactive-saturated
phase boundaries join at a bicritical point where the universal
behavior is in the even branching annihilating random walk class.  We
find the crossover exponent from bicritical to critical behavior and a
new exponent associated with the bicritical interface dynamics.

\end{abstract}


\maketitle
\begin{multicols}{2}
\narrowtext

Nonequilibrium models with many degrees of freedom whose dynamics
violate detailed balance arise in studies of biological populations,
heterogeneous catalysis, fluid turbulence, and elsewhere.  The
macroscopic behavior of these models can be much richer than that of
systems in thermal equilibrium, showing organized macroscopic spatial
and temporal structures like pulses or waves, 
and even spatiotemporal chaos. 
Even the steady state behavior can be far more complicated,
involving for example generic scale invariance.
Like their equilibrium cousins, nonequilibrium
systems at continuous transitions between steady states
show universal behavior that is insensitive to microscopic details and
depends only on properties such as symmetries and conservation laws.

One of the most common continuous phase transitions in nonequilibrium
models is a transition to an absorbing, noiseless steady
state\cite{M+D}, the term absorbing indicating the state cannot be left
once it is reached.  Examples of this include directed percolation
(DP)\cite{DP1,DP2}, the contact process \cite{CP}, the dimer poisoning
transition in the ZGB model\cite{ZGB} for the catalytic oxidation of
CO, auto-catalytic reaction models\cite{ABW}, and branching
annihilating random walks with odd numbers of
offspring\cite{BAW2,BAW3}.  Both renormalization group
calculations\cite{DP1,RFT} and Monte Carlo
simulations\cite{DP2,CP,ZGB,ABW,BAW2,BAW3} show that these models form
a single universality class for a purely nonequilibrium model with no
internal symmetry in the order parameter.

Recently, a number of models with continuous adsorbing transitions
in a universality class distinct from directed percolation
have been studied.  These models include
the probabilistic cellular automata models studied of Grassberger {\em
et al.\/}\cite{PCA}, certain kinetic Ising models \cite{NKI}, the
interacting monomer-dimer model\cite{IMD1,IMD2}, and branching 
annihilating random walks with an even number of offspring
(BAWe)\cite{BAW2,BAWe}.  All of these models except for the BAWe
have two equivalent absorbing states indicating the importance
of symmetry of the adsorbing state to the universality class.
However, the universal behavior of this new class
is apparently controlled by a dynamical conservation law.  
If\cite{PCA} the important dynamical
variables in this class are defects represented by the walkers in the
BAWe model and the walls between different saturated domains in the
other models,  the models have ``defect parity'' conservation law where
the number of defects is conserved modulo 2.  Recent field theoretic
work confirms this viewpoint\cite{CardyRG}.

We study here a simple catalytic surface reaction model with three
different equivalent monomer species.  This model could represent
either a system with three different chemical species or an
auto-catalytic reaction system in which one chemical species can adsorb
on three different types of surface sites.  This model has adsorbing
transitions to both one and two equivalent noiseless states, and
therefore is a good model to study the role of symmetry in adsorbing
phase transitions.  The model has two fundamental processes:  (a)
monomer adsorption onto sites of a substrate, and (b) the annihilation
reaction of two dissimilar monomers adsorbed on nearest-neighbor sites
of the substrate.  Here we consider the model only in the adsorption
controlled limit where process (b) occurs instantaneously.  We present
here results only for the one dimensional version of the model.

\begin{figure} 
\centerline{\psfig{file=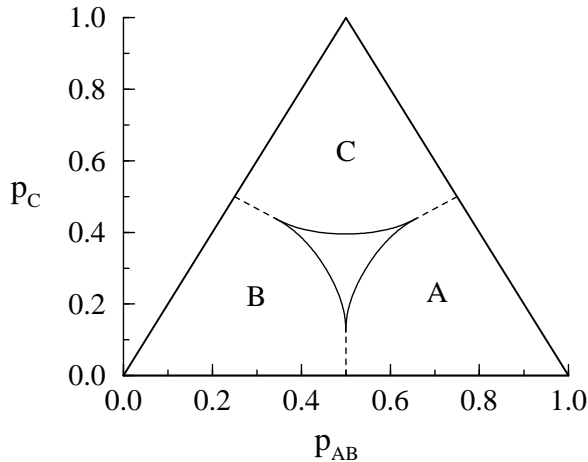,width=3.2in}} 
\caption{ Phase diagram showing three saturated phases (indicated by the
letters), and a reactive phase (the unlabeled center region). Solid
lines indicate continuous transitions. Dashed lines indicate
first-order transitions.  } 
\label{fig1} 
\end{figure}

Calling the monomer species $A$, $B$ and $C$, the parameters in the
model are then the relative adsorption rates of the different monomer
species $p_A$, $p_B$, and $p_C$, such that $p_A + p_B + p_C = 1$.
Using static Monte Carlo simulations to find the steady state, and
dynamical Monte Carlo studies described below, we find the ternary
phase diagram for the model is shown in Fig.~\ref{fig1}, where the
horizontal axis corresponds to the relative adsorption rate of $A$ and
$B$ monomers $p_{AB} = p_A/(p_A + p_B)$.  There are noiseless phases
where one monomer species saturates the chain occupying the corners of
the phase diagram and a reactive steady state in the center.  There are
continuous phase transitions from the reactive phase to the saturated
phases.  The monomer densities undergo discontinuous, first-order,
transitions from one saturated state to another.  The points where the
reactive phase and two saturated phases meet are {\em bicritical}
points\cite{BCP} where two lines of continuous transitions meet a line
of first order transitions.  We have also constructed a mean field
theory of the model, to be presented elsewhere\cite{intheworks},
following the methods of Dickman\cite{DickmanMFT}.  One unusual feature
of the mean field phase diagram is that the bicritical points lie on 
the edge of the phase diagram if the correlations are correct up to 
single sites or even nearest neighbor pairs.  Only when the 
correlations up to triplets of adjacent sites are included does 
the bicritical point appear inside the phase diagram, 
indicating the importance of reproducing the correlations induced 
by large domains of a single saturated phase.  

We have used dynamical Monte Carlo simulations to investigate the
universality classes of the continuous transitions and bicritical
points, the critical dynamics of interfaces between the two symmetric
saturated states at the bicritical points, 
the crossover behavior near the bicritical point,
and the first-order lines.  We use two forms of ``epidemic''
analysis\cite{BAW3,BAWe,DMCS} following the evolution of an initial
condition chosen close to the saturated state.  In the first form
(defect dynamics) we use an initial condition consisting of a single
vacancy in a saturated phase.  The second (interface dynamics) starts
from two different saturated phases separated by a single vacancy.

From the simulations we find the probability $P(t)$ the system
does not fall into the saturated state in $t$ time steps, the average
number of vacancies per run $\langle{n}_V(t)\rangle$, and the typical
size of the defect or interface per surviving run
$\langle{R^2}(t)\rangle$. At a continuous phase transition as
$t\rightarrow\infty$ it is expected that they obey power law behavior
\begin{equation}
P(t) \sim t^{-\delta} , \qquad
\langle{n}_V(t)\rangle \sim t^{\eta} , \qquad
\langle{R^2}(t)\rangle \sim t^{z} .
\label{quantities}
\end{equation}
Precise estimates of the location of the critical point and of the
exponents can be made by examining the local slopes of the curves on a
log-log plot and extrapolating to infinite times.  Then, the effective
exponent $\delta(t)$ is
\begin{equation}
-\delta(t) = \{ \ln\left[ P(t)/P(t/b) \right] / \ln b \},
\label{localslopes}
\end{equation}
$\eta(t)$ and $z(t)$ being defined analogously. Plotting the local
slopes versus $t^{-1}$ allows us to determine both the exponent and the
critical point accurately.  At the critical point the local slope will
extrapolate to a constant as $t^{-1}\to0$ with a scaling correction
linear in $t^{-1}$\cite{ctsnote}.  Data taken away from the critical
point will have local slopes that curve away from the critical point
value as $t^{-1}\to0$.

\begin{figure}
\centerline{\psfig{file=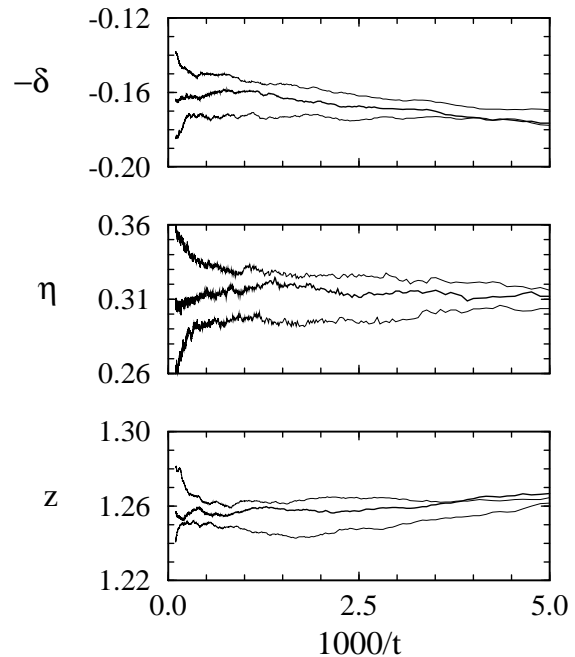,width=3.2in}}
\caption{
Effective exponents using Eq.~(\protect\ref{localslopes}) with
$b=5$ for the defect dynamics
near the critical point at $p_{AB}=0.5$ on the line where the
C poisoned phase meets the reactive phase.
From top to bottom, the 3 curves in each panel correspond to
$p_C = 0.3955$, 0.39575, and 0.3960, with the middle curve
corresponding to the critical point.
}
\label{figdp}
\end{figure}

Figure~\ref{figdp} shows the effective exponents of the three dynamic
quantities near the phase transition to the $C$ saturated phase at
$p_{AB}=0.5$.  Using $10^5$ independent runs of up to $10^4$ time steps
at each parameter value, we found a critical $C$ monomer adsorption
rate of $\tilde{p}_C = 0.39575(10)$, and the critical exponents are
$\delta = 0.16(1)$, $\eta = 0.31(1)$, and $z = 1.255(15)$.  These
values are consistent with our expectation that this transition should
be in the DP universality class, for which the exponents are $\delta =
0.1596(4)$, $\eta = 0.3137(10)$, and $z = 1.2660(14)$\cite{DPexp}.  We
found similar exponents for the adsorbing transition at a number of
other points along the lines separating the reactive phase and the
saturated states, indicating the transition between the reactive phase
and any single saturated phase is always in the DP universality class.

\begin{figure}
\centerline{\psfig{file=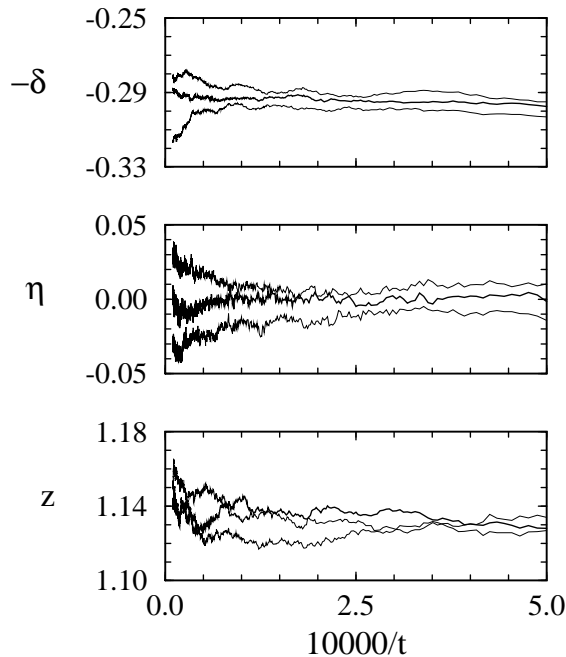,width=3.2in}}
\caption{
Effective exponents for the defect dynamics
near the bicritical point where the
A and B poisoned phases meet the reactive phase
as defined in (\protect\ref{localslopes}) with $b=5$.
From bottom to top, the 3 curves in each panel correspond to
$p_C = 0.121$, 0.122, and 0.123, with the middle line
corresponding to the bicritical point.
}
\label{fig2}
\end{figure}

The same kind of analysis at the bicritical point at $p_{AB}=0.5$,
using an initial condition of a vacancy in an A-saturated phase, yields
a bicritical point at $p_C = p_C^{*} = 0.122(1)$, and very different
exponents, given the presence of two-symmetry equivalent saturated
phases.  From $5\times10^5$ runs of up to $10^5$ time steps we found
the local slope data shown in Fig.~\ref{fig2}, yielding values of
$\delta= 0.29(1)$, $\eta = 0.00(1)$, and $z = 1.150(15)$.  These values
indicate that the bicritical behavior falls in the BAWe universality
class, for which $\delta = 0.285(2)$, $\eta = 0.000(1)$, and $z =
1.141(2)$\cite{BAWe}.

For $p_C < p^{*}_C$ along the A-B coexistence line, a similar analysis
yields the sub-critical dynamic exponents 
$\delta \approx 0.5$, $\eta \approx -0.5$, and $z \approx 1$.
These exponents also describe the two species version of the
model which can be mapped onto the well known problem of the $T=0$ 
one-dimensional kinetic Ising model for which these values are known
exactly\cite{zeroTIsing}. 

To further analyze the importance of competition in the growth of two
equivalent saturated phases at the bicritical point we also studied the
dynamics of an interface between those two phases.  Starting with a
single vacancy between the two domains, we used two different methods
to analyze the behavior of the interface.  Since there must always be
at least one vacancy between two different saturated phases, in the
first method we ignore the survival probability $P(t)$ and take
$\delta\equiv0$.  We then measure the number of vacancies in the
interface $\langle n(t)\rangle \propto t^{\eta}$ and average size of
the interface $\langle R^2(t)\rangle \propto t^{z}$.  From
$5\times10^4$ independent runs at the bicritical point, each lasting
$10^5$ time steps, we found the other exponents to be $\eta=0.285(10)$
and $z=1.14(2)$.  This type of interface dynamics has been used to
study the properties of critical interfaces in other models in the BAWe
class, where similar results for $\eta$ and $z$ were
obtained\cite{IMD2,BAWe}.

\begin{figure}
\centerline{\psfig{file=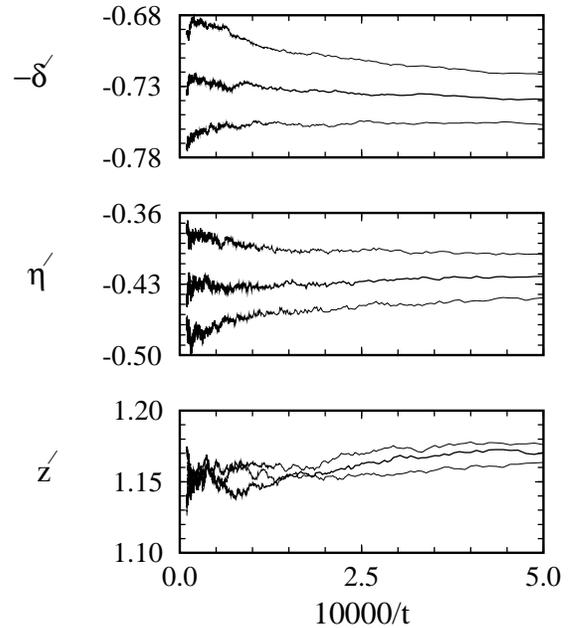,width=3.2in}}
\caption{
Effective exponents, as in Fig.~\protect\ref{fig2},
for the second type of interface dynamics
near the bicritical point where the
A and B poisoned phases meet the reactive phase.
}
\label{fig3}
\end{figure}

In the second type of interface dynamics simulations, which has not
been studied before, the simulation is stopped if the interface between
the domains has ``collapsed'' back to one vacant site.  We introduce a
probability of avoiding a collapse $P(t)\propto t^{-\delta'}$ and
corresponding vacancy concentrations $\langle n(t)\rangle \propto
t^{\eta'}$ and $\langle R^{2}(t)\rangle \propto t^{z'}$.
Figure~\ref{fig3} shows results from $10^7$ independent runs each
lasting up to $10^5$ time steps.  We find values of $\delta' =
0.73(2)$, $\eta'=-0.43(2)$ and $z'=1.15(2)$.

Note the value of the dynamic exponent $z$ or $z'$, which measures the
size of the active region during surviving runs, is the same in both
types of interface dynamics simulations as that measured for the defect
dynamics.  Furthermore, although the exponents $\delta$ and $\eta$ are
different in the three cases, their sum $\delta + \eta$ (or
$\delta'+\eta'$), which governs the time evolution of the number of
vacancies in just the surviving runs, seems to be the same.  This
indicates a universal nature of the critical spreading of the active
region for models with two symmetric adsorbing states which is
independent of whether defect or interface dynamics is being
considered.  A similar result holds for some one-dimensional systems
with infinitely many adsorbing states\cite{ManyAS}.

Assuming this conjecture is true, it should be noted that simulations
using the first type of interface dynamics, where $\delta \equiv 0$,
yield no information beyond that obtainable from simulations employing
defect dynamics.  However, simulations using the second type of
interface dynamics measure an independent dynamic exponent $\delta'$
which we expect to be a universal number. It would be interesting to
measure this exponent for other models in the BAWe class.

\begin{figure}
\centerline{\psfig{file=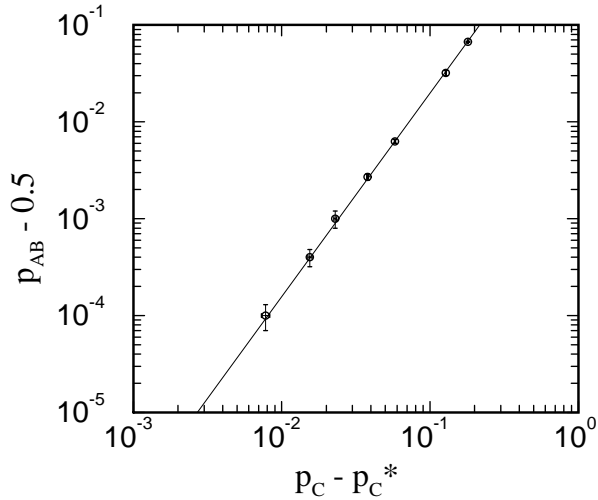,width=3.2in}}
\caption{
Location of the critical line as a function of distance from the
bicritical point. The data falls on a line with a slope
corresponding to the crossover exponent $\phi = 2.1\pm0.1$.
}
\label{fig4}
\end{figure}

Finally, we measured the crossover from bicritical to critical
behavior.  Near the bicritical point where the A and B poisoned phases
meet, the boundary of the reactive region is expected to behave as
$(p_{AB} - 0.5) \propto (p_C - p^{*}_C)^{\phi}$, where $\phi$ is the
crossover exponent\cite{BCP}.  We used the dynamical simulation method
to accurately determine the location of the DP phase boundary between
the reactive phase and the $A$ saturated phase near the bicritical
point.  From the log-log plot of $p_{AB}-0.5$ versus $p_C-p^{*}_C$
shown in Fig.~\ref{fig4} we find $\phi=2.1\pm0.1$.

We have introduced a convenient model to study the role of symmetry in
the critical dynamics of adsorbing phase transitions. We have shown
that the universality class of the transition changes from DP to BAWe
when the symmetry of the adsorbing state is increased from one to two
equivalent noiseless states.  Furthermore, we have shown that having a
symmetry in the adsorbing states introduces a richness into the
dynamics that is not possible if there is a unique adsorbing state.  In
particular, the critical dynamics of the interfaces between two
different adsorbing states shows a sensitivity to how the dynamics is
defined, and the survival probability of  fluctuations in the size of
the interface from its smallest value is described by a new universal
exponent $\delta'$.  However, the critical spreading of the reactive
region, be it a defect in a single phase or a domain wall between
phases, appears to be insensitive to the choice of initial conditions.
This appears to result from the fact that large reactive regions are
insensitive to whether the reactive regions are bounded by the same or
different saturated phases.

This work was supported by the National Science Foundation
under Grant No.~DMR--9408634.

\end{multicols}


\begin{references}

\bibitem{M+D}
J. Marro and R. Dickman,
{\em Nonequilibrium phase transitions in lattice models}
(Cambridge Univ. Press, 1996).

\bibitem{DP1}
H. K. Janssen, Z. Phys. B {\bf 42}, 151 (1981).

\bibitem{DP2}
P. Grassberger, Z. Phys. B {\bf 47}, 365 (1982).

\bibitem{CP}
T. E. Harris, Ann. Prob. {\bf 2}, 969 (1974).

\bibitem{ZGB}
R. M. Ziff, E. Gulari, and Y. Barshad,
Phys. Rev. Lett. {\bf 56}, 2553 (1986);
G. Grinstein, Z.-W. Lai, and D. A. Browne,
Phys. Rev. A {\bf 40}, 4820 (1989);
I. Jensen, H. C. Fogedby, and R. Dickman,
{\em ibid.} {\bf 41}, 3411 (1990).

\bibitem{ABW}
T. Aukrust, D. A. Browne, and I. Webman,
Phys. Rev. A {\bf 41}, 5294 (1990).

\bibitem{BAW2}
H. Takayasu and A. Yu. Tretyakov,
Phys. Rev. Lett. {\bf 68}, 3060 (1992).

\bibitem{BAW3}
I. Jensen, J. Phys. A {\bf 26}, 3921 (1993).

\bibitem{RFT}
J. L. Cardy and R. L. Sugar,
J. Phys. A {\bf 13}, L423 (1980).

\bibitem{PCA} P. Grassberger, F. Krause, and T. von der Twer,
J. Phys. A {\bf 17}, L105 (1984);
P. Grassberger, J. Phys. A {\bf 22}, L1103 (1989).

\bibitem{NKI}
N. Menyh\'{a}rd, J. Phys. A {\bf 27}, 6139 (1994);
N. Menyh\'{a}rd and G. \'{O}dor, J. Phys. A {\bf 27}, 6139 (1994).

\bibitem{IMD1}
M. H. Kim, and H. Park,
Phys. Rev. Lett. {\bf 73}, 2579 (1994);
H. Park, and H. Park,
Physica A {\bf 221}, 97 (1995).

\bibitem{IMD2}
H. Park, M. H. Kim, and H. Park,
Phys. Rev. E {\bf 52}, 5664 (1995).

\bibitem{BAWe}
I. Jensen,
Phys. Rev. E {\bf 50}, 3623 (1994).

\bibitem{CardyRG}
J. Cardy and U. Ta\"uber, preprint.

\bibitem{BCP}
M.E. Fisher and D.R. Nelson,
Phys. Rev. Lett. {\bf 32}, 1350
(1974).

\bibitem{intheworks} K. E. Bassler and D. A. Browne, to be published.

\bibitem{DickmanMFT} R. Dickman, Phys.\ Rev.\ A {\bf 34}, 4626 (1986).

\bibitem{DMCS}
P. Grassberger, J. Phys. A {\bf 22}, 3673 (1989);
P. Grassberger and A. de la Torre, Ann. Phys. (New York) {\bf 122}, 373 (1979).

\bibitem{ctsnote}
Actually, this statement is known to be true only
for directed percolation \protect\cite{DP2}, but it seems to also be
true\protect\cite{BAWe} for the BAWe universality class.

\bibitem{DPexp}
I. Jensen and R. Dickman, J. Stat. Phys. {\bf 71}, 89 (1993).

\bibitem{zeroTIsing}
A.A. Lushnikov, Phys. Lett. {\bf 120A}, 135 (1987);
J.L. Spouge, Phys. Rev. Lett. {\bf 60} 871 (1988); 
J.G. Amar and F. Family, Phys. Rev. A {\bf 41}, 3258 (1990).


\bibitem{ManyAS}
J.F.F. Mendes, R. Dickman, M. Henkel, and M.C. Marques,
J. Phys. A {\bf 27}, 3019 (1994).

\end{references}
\end{document}